\documentclass[12pt]{article}
\newcommand{\be}{\begin{equation}}
\newcommand{\ee}{\end{equation}}
\newcommand{\ba}{\begin{eqnarray}}
\newcommand{\ea}{\end{eqnarray}}

\usepackage{amssymb}
\usepackage{amsmath}
\begin{document}
\begin{center}
{\bf TWO-DIMENSIONAL SUPERSYMMETRY: FROM SUSY QUANTUM
MECHANICS TO INTEGRABLE CLASSICAL MODELS}\\
\vspace{0.4cm} {\large\bf M. V. Ioffe$^{1,2,}$\footnote{E-mail:
m.ioffe@pobox.spbu.ru}, J. Mateos Guilarte$^{3,}$\footnote{E-mail:
guilarte@usal.es},
P. A. Valinevich$^{1,}$\footnote{E-mail: pavel@PV7784.spb.edu}}\\
\vspace{0.2cm}
$^1$ Department of Theoretical Physics, Sankt-Petersburg State University,\\
198504 Sankt-Petersburg, Russia\\
$^2$ Departamento de Fisica Teorica, Atomica y Optica, Universidad
de Valladolid,
47071 Valladolid, Spain\\
$^3$ Departamento de Fisica Fundamental and IUFFyM, Facultad de
Sciencias, Universidad de Salamanca, 37008 Salamanca, Spain
\end{center}
\vspace{0.2cm} \hspace*{0.5in} \hspace*{0.5in}
\begin{minipage}{5.0in}
{\small Two known 2-dim SUSY quantum mechanical constructions -
the direct generalization of SUSY with first-order supercharges
and Higher order SUSY with second order supercharges - are
combined for a class of 2-dim quantum models, which {\it are not
amenable} to separation of variables. The appropriate classical
limit of quantum systems allows us to construct SUSY-extensions of
original classical scalar Hamiltonians. Special emphasis is placed
on the symmetry properties of the models thus obtained - the
explicit expressions of quantum symmetry operators and of
classical integrals of motion are given for all (scalar and
matrix) components of SUSY-extensions. Using Grassmanian
variables, the symmetry operators and classical integrals of
motion are written in a unique form for the whole
Superhamiltonian. The links of the approach to the classical
Hamilton-Jacobi method for related "flipped" potentials are
established.
\\
\vspace*{0.1cm} PACS numbers: 03.65.-w, 03.65.Fd, 11.30.Pb }
\end{minipage}
\vspace*{0.2cm}
\section*{1. Introduction}
\vspace*{0.5cm} \hspace*{3ex} Supersymmetric Quantum Mechanics
(SUSY QM) \cite{cooper} is a new framework for analyzing
non-relativistic quantum problems. In particular, it helps to
investigate the spectral properties of different quantum models as
well as to generate new systems with given spectral
characteristics (quantum design).
\par Much less attention has been paid to SUSY QM as a tool to
study individual hidden symmetries of the superpartner
Hamiltonians. This problem is reasonable for considering either
one-dimensional quantum systems with internal degrees of freedom
(with matrix potentials) or systems of higher spatial
dimensionality. Only for these classes of systems may the symmetry
operators, which are in involution with the Hamiltonian (and are
independent of it), exist. Thus, the quantum models in
one-dimensional SUSY QM with matrix potentials and the
higher-dimensional (in particular, two-dimensional) models with
scalar and/or matrix potentials are extremely attractive.
\par Regarding SUSY QM systems with matrix potentials, we refer
to papers \cite{matrix}. The SUSY QM systems with an arbitrary ($d
> 1$) dimensionality of space\footnote{Instead of dimensionality
of the space $d$ can be also interpreted as a number of particles,
for example in Calogero-like models of interacting $N$ particles
on a line \cite{calogero}.} were constructed and investigated in
\cite{2d}, \cite{abei}. Such models include both the scalar and
matrix components of the Superhamiltonian. The latter are
interesting either for a description of interacting
non-relativistic particles with spin \cite{scattering},
\cite{neelov}, or for developing supersymmetric quantum field
theory on a spatial lattice \cite{wipf}. The appearance both of
scalar and matrix Hamiltonians in a unique Superhamiltonian
provides an opportunity to consider (starting from a given scalar
Schr\"odinger operator) SUSY extensions that correspond to the
systems with internal degrees of freedom. Some SUSY extensions of
this type were considered in \cite{neelov1} (Calogero-like models
of $N$ particles on a line) and in \cite{wipf2} (Coulomb potential
in $d$ dimensions).
\par Among multidimensional SUSY QM models,
those most developed are the two-dimensional ones. Namely,
precisely for these systems second-order supercharges were used to
build the higher-order deformation of SUSY algebra \cite{david1},
\cite{david2}, \cite{ioffe}. In the framework of this HSUSY QM
generalization of conventional Witten's
superalgebra\footnote{One-dimensional HSUSY QM was investigated in
detail in \cite{acdi}, \cite{highod}. }
 one can
avoid the appearance of any matrix components of the
Superhamiltonian, so that two {\it scalar} two-dimensional
Schr\"odinger operators are intertwined by second-order
supercharges. As a by-product of this construction, each of the
intertwined Hamiltonians obeys the hidden symmetry: the
differential operators of fourth order in derivatives exist, which
are not reducible to the Hamiltonian and commute with the
Hamiltonian, \cite{david1}, \cite{david2}, \cite{newmet},
\cite{ioffe}. In the two-dimensional context this means the
complete integrability of the system.
\par Another direction in which to investigate SUSY QM models involves
their connections with classical counterparts. Initially, the
arbitrary-space-dimensionality generalization \cite{abei} of SUSY
QM, mentioned above, was obtained by canonical quantization of a
suitably chosen multi-dimensional system of Classical Mechanics.
Then the quasi-classical limit of some supersymmetrical quantum
models investigated afforded new insight into the properties of
the classical models obtained. In the one-dimensional case, this
limit led to a new SWKB quantization rule \cite{swkb}, which
turned out to be more useful than the standard WKB rule. In the
two-dimensional case, the quasi-classical limit provided an
alternative effective method \cite{david4} for the construction of
integrable systems in Classical Mechanics, which essentially
enlarges the list of such models. The SUSY QM approach also
provides new interesting links with the well-known Hamilton-Jacobi
equation in Classical Mechanics \cite{Aai1}.
\par In the present paper we shall combine both known two-dimensional
SUSY QM constructions - the direct generalization of SUSY with
first-order supercharges and HSUSY with second-order supercharges
- in order to investigate the symmetry properties of the models
obtained, both at quantum and classical level. The paper is
organized as follows. In Section 2, known results about
two-dimensional SUSY QM models and their connections with
classical models, necessary for the original part of the paper,
are briefly summarized. In Section 3, the particular case of
two-dimensional models for second-order supercharges with
intermediate twists are studied: the particular models of
generalized Morse and P\"oschl-Teller potentials are presented
within this common framework. In Section 4, the integrability of
these models is extended onto their matrix - both quantum and
classical - superpartners. In Section 5, the links with classical
Hamilton-Jacobi equation are considered. \vspace*{0.2cm}
\section*{2. Two-dimensional SUSY Quantum Mechanics}
\vspace*{0.2cm}
\subsection*{2.1. The representation of SUSY algebra with first-order supercharges}
\vspace*{0.5cm} \hspace*{3ex}
\par In the two-dimensional case $\vec x = (x_1, x_2)$, the direct (i.e. of first order in
derivatives) generalization of SUSY QM satisfies \cite{2d},
\cite{abei} the conventional Witten's \cite{witten} SUSY algebra
\begin{equation} \{ \hat{Q}^+, \hat{Q}^- \} = \hbar\hat{H};\qquad \{
\hat{Q}^+, \hat{Q}^+ \} = \{ \hat{Q}^-, \hat{Q}^- \} = 0;\qquad [
\hat{Q}^\pm, \hat{H} ] = 0.\label{alg}
\end{equation}
by the 4$\times$4 matrix operators:
\begin{equation} \hat H = \left(
\begin{array}{ccc}H^{(0)}(\vec x)&0&0\\ 0& H^{(1)}_{ik}(\vec
x)&0\\0&0&H^{(2)}(\vec x)
\end{array} \right);\quad i,k=1,2;\quad \hat Q^+ = (\hat Q^-)^{\dagger} =
\hbar^{1/2}\left(
\begin{array}{cccc}0&0&0&0\\
q_1^-&0&0&0\\q_2^-&0&0&0\\0&s_1^+&s_2^+&0
\end{array} \right),\label{suham}
\end{equation}
where two scalar Hamiltonians $H^{(0)}, H^{(2)}$ and one
2$\times$2 matrix Hamiltonian $H^{(1)}_{ik}$ of the Schr\"odinger
type take on a quasi-factorized form:
\begin{eqnarray}
H^{(0)}&=&q_l^+q_l^- = -\hbar^2\partial_l^2 + V^{(0)}({\vec x}) =
-\hbar^2\partial_l^2 + \Bigl(\partial_l\chi (\vec x)\Bigr)^2 -
\hbar\partial_l^2\chi (\vec x);\,\, \partial_l^2 \equiv
\partial_1^2 + \partial_2^2; \nonumber\\
H^{(2)}&=&s_l^+s_l^-=-\hbar^2\partial_l^2 + V^{(2)}({\vec x}) =
-\hbar^2\partial_l^2 + \Bigl(\partial_l\chi (\vec x)\Bigr)^2 +
\hbar\partial_l^2\chi (\vec x);\label{hams}\\
H^{(1)}_{ik}&=&q_i^-q_k^+ + s_i^-s_k^+ = -\hbar^2\delta_{ik}
\partial_l^2 + \delta_{ik}\Bigl(\bigl(\partial_l\chi (\vec x)\bigr)^2 -
\hbar\partial_l^2\chi (\vec x)\Bigr) + 2\hbar\partial_i\partial_k
\chi (\vec x),\nonumber
\end{eqnarray}
the components of the supercharges being of first order in
derivatives:
\begin{equation}
q_l^{\pm}\equiv \mp\hbar\partial_l + (\partial_l \chi (\vec
x));\quad s_l^{\pm}\equiv \epsilon_{lk}q_k^{\mp},\label{2p}
\end{equation}
where $\partial_i \equiv \partial/\partial x_i$ and summation over
repeated indices is assumed. The Planck constant was restored in
formulas (\ref{suham}) - (\ref{2p}), and the ({\it normalizable or
unnormalizable}) zero-energy wave functions of the scalar
Hamiltonians $H^{(0),(2)}$ are now written as
$\exp{(\pm\chi/\hbar)}$.
\par The quasi-factorization in (\ref{hams})
ensures that the last equation in (\ref{alg}) holds, and leads to
the following intertwining relations for the components of the
Superhamiltonian (\ref{suham}):
\begin{equation}
H^{(0)}q_i^+=q_k^+H^{(1)}_{ki};\quad H^{(1)}_{ik}q_k^- =
q_i^-H^{(0)}; \quad H^{(1)}_{ik}s_k^-=s_i^-H^{(2)} ; \quad
H^{(2)}s_i^+ = s_k^+H^{(1)}_{ki}.\label{2dintw}
\end{equation}
These relations play the main role in the SUSY QM approach and, in
particular, they lead to the connections between the spectrum of
the matrix Hamiltonian and the spectra of two scalar ones. We
remark that $H^{(0)}$ and $H^{(2)}$ {\it are not} intertwined with
each other and {\it are not} (in general) isospectral since
$q^+_k\cdot s^-_k \equiv 0$. \vspace*{0.2cm}
\subsection*{2.2. Second-order supercharges}
\vspace*{0.5cm} \hspace*{3ex} An alternative opportunity to
include two-dimensional scalar Hamiltonians in the SUSY QM
framework is based on the supercharges of second order in
derivatives \cite{david1}, \cite{david2}, \cite{ioffe}:
\begin{equation}
Q^+ = (Q^-)^\dag = g_{ik}(\vec{x})\hbar^2\partial_i\partial_k +C_i
(\vec{x})\hbar\partial_i + B(\vec{x}) \label{Q}
\end{equation}
where $g_{ik}, C_i, B$ are arbitrary real functions. In this case,
two scalar Hamiltonians $H^{(0)}$, $\widetilde{H}^{(0)}$ are
intertwined directly without any (matrix) intermediary:
\begin{equation}
\widetilde{H}^{(0)}(\vec{x})Q^+ = Q^+H^{(0)}(\vec{x});\qquad
H^{(0)}(\vec{x})Q^- = Q^-\widetilde{H}^{(0)}(\vec{x}).
\label{intw2d}
\end{equation}
Although no method to find the general solution of the
intertwining relations (\ref{intw2d}) has been proposed, a certain
number of models obeying these relations  have been found for the
cases of hyperbolic (Lorentz) and degenerate metric $g_{ik}.$ One
very important specific property of all these models, which
follows from the intertwining relations (\ref{intw2d}), is their
integrability. Indeed, both Hamiltonians possess the symmetry
operators $R^{(0)}, \widetilde{R}^{(0)}$ of fourth order in
derivatives \cite{david1}, \cite{david2}: \be [R^{(0)}, H^{(0)}] =
0;\quad [\widetilde{R}^{(0)}, \widetilde{H}^{(0)}] = 0;\qquad
R^{(0)} = Q^-Q^+; \qquad \widetilde{R}^{(0)} = Q^+Q^-,
\label{symm} \ee which are not, in general\footnote{It has been
proved \cite{david2} that only for Laplacian (elliptic) metric
$g_{ik}=\delta_{ik}$ can the symmetry operators be reduced to
second-order operators, and the corresponding Hamiltonians
$H^{(0)}, \widetilde{H}^{(0)}$ allow the separation of
variables.}, polynomials of $H^{(0)}, \widetilde{H}^{(0)}$.
\par For the case  of Lorentz
(hyperbolic) metric $g_{ik} = diag(1,-1)$ \cite{david1} -
\cite{david4}, the intertwining relations (\ref{intw2d}) can be
rewritten in a reduced form:
\begin{equation}
\partial_-(C_-F)  = -
\partial_+(C_+F), \label{2odmain}
\end{equation}
where $x_\pm = (x_1 \pm x_2)/\sqrt{2}$; functions $C_{1,2}$ were
found to satisfy $C_\pm \equiv C_1 \mp C_2 \equiv
C_\pm(\sqrt{2}x_\pm),$ and $F(\vec{x})= F_1(2x_1)+ F_2(2x_2)$.
Thus, the potentials $V^{(0),(1)}$ and the supercharges $Q^{\pm}$
are expressed in terms of the functions $C_\pm(\sqrt{2}x_\pm)$ and
$F_1(2x_1), F_2(2x_2)$:
\begin{eqnarray}
V^{(0)}, \widetilde{V}^{(0)} & = &
\mp\frac{1}{2}\hbar\Bigl(C'_+(\sqrt{2}x_+) +
C'_-(\sqrt{2}x_-)\Bigr) + \frac{1}{8}\Bigl(C_+^2(\sqrt{2}x_+) +
C_-^2(\sqrt{2}x_-)\Bigr)  +\nonumber\\
&&+\frac{1}{4}\Bigl(F_2(2x_2) - F_1(2x_1)\Bigr); \label{ve2od}\\
Q^+ & = & \hbar^2(\partial_1^2 - \partial_2^2) +
C_1\hbar\partial_1 + C_2\hbar\partial_2 + B;\label{quplus}\\ B & =
& \frac{1}{4}\Bigl(C_+(\sqrt{2}x_+)C_-(\sqrt{2}x_-) + F_1(2x_1) +
F_2(2x_2)\Bigr),\label{be2od}
\end{eqnarray}
where the prime denotes the derivative of function with respect to
its argument. A set of particular solutions of (\ref{2odmain}) was
obtained in \cite{david2}, \cite{david4}, \cite{iv}.
\vspace*{0.2cm}
\section*{3. Two-dimensional models with twisted reducibility of supercharges}
\vspace*{0.5cm} \hspace*{3ex} In the previous Section it was shown
that two different constructions with very different properties
exist in two-dimensional SUSY QM. The first one (Subsection 2.1.)
includes two scalar Hamiltonians $H^{(0)},\,\, H^{(2)}$ (only one
of them has normalizable\footnote{For the case of supersymmetry
not broken spontaneously.} zero-energy ground state wave function
$\Psi_0(\vec x) = \exp (\pm\frac{\chi (\vec x)}{\hbar})$) and
their $2\times2$ matrix partner $H^{(1)}_{ik}.$ The second one
(Subsection 2.2.) contains only the scalar Hamiltonians
$H^{(0)},\,\, \widetilde{H}^{(0)}$ with no information about their
ground-state energy in advance, and both $H^{(0)}$ and
$\widetilde{H}^{(0)}$ {\it a priori} (by construction) obey the
important property of integrability with the symmetry operators
$R^{(0)},\,\widetilde{R}^{(0)}$  of fourth order in derivatives
(see (\ref{symm})). The natural idea is to unite all the above
tempting properties by combining these two constructions, i.e. by
identifying the original Hamiltonian $H^{(0)}$ as the same in both
approaches. More precisely: let $H^{(0)}$ of the form (\ref{hams})
have the superpartners $H^{(1)}_{ik}$ and $H^{(2)}$ in the
first-order scheme, and {\it at the same time} the superpartner
$\widetilde{H}^{(0)}$ in the second-order scheme. It is known
\cite{david2} that the simplest, {\it reducible or
quasi-factorizable}, form of the second order supercharges
$Q^{+}=(Q^-)^{\dagger}=q_i^{+}\tilde q_i^{-}$, which is suitable
for the construction described, leads to the $R-$separation of
variables, and therefore it is not considered here. All other
models (excluding the case of elliptic metric $g_{ik}=\delta_{ik}$
in $Q^{\pm}$) have been proved \cite{david2} {\it not} to be
amenable to separation of variables; they have nontrivial
fourth-order symmetry operators (\ref{symm}). The main idea to
achieve the identification of $H^{(0)}$ in the two approaches is
to consider a class of models with second-order supercharges,
which are quasi-factorizable, {\bf but} with an intermediate {\it
twist} transformation (see also \cite{iv}): \be
Q^-=(Q^+)^{\dagger}=q_i^+U_{ik}\tilde q^-_k ,\label{twist} \ee
where $U_{ik}$ is a constant unitary matrix, $q^{\pm}$ were
defined in (\ref{2p}), and $$ \tilde q^{\pm}_k \equiv \mp
\hbar\partial_k + (\partial_k\tilde\chi (\vec x))
$$ with some new superpotential $\tilde\chi .$ Such a generalization
of the notion of reducibility (we shall call it {\it twisted
reducibility}) is somehow reminiscent of the "gluing with shift"
recipe in one-dimensional scalar \cite{acdi} and matrix
\cite{matrix} HSUSY QM. The intertwining relations (\ref{intw2d})
with supercharges (\ref{twist}) and the general expression for
matrix $U_{ik} :$
$$
U = \alpha_0\sigma_0 +
i\overrightarrow{\alpha}\overrightarrow{\sigma}; \quad \alpha_0^2
+ \overrightarrow{\alpha}^2 = 1; \quad \alpha_0, \alpha_i \in
\bf{R},
$$
($\sigma_i$ are the Pauli matrices and $\sigma_0$ is the unit
matrix) give the system of four linear and one nonlinear equations
for two functions $\chi_\pm = \frac{1}{2}(\chi \pm \tilde{\chi}):$
\begin{eqnarray}
\alpha_3\Box\chi_- +2\alpha_1\partial_1\partial_2\chi_- = 0;
\qquad \alpha_1\Box\chi_+ - 2\alpha_3\partial_1\partial_2\chi_+
&= 0;\label{lin1}\\
\alpha_2\Box\chi_+ - 2\alpha_0\partial_1\partial_2\chi_- = 0;
\qquad \alpha_0\Box\chi_- + 2\alpha_2\partial_1\partial_2\chi_+ &=
0; \label{lin2}
\end{eqnarray}
\begin{equation}
(\partial_k \chi_-)(\partial_k\chi_+) = 0, \label{nonlin}
\end{equation}
where $\Box \equiv \partial_1^2 - \partial_2^2$. Precisely the last
equation (\ref{nonlin}) is obviously most difficult to solve. Both
the solutions of linear partial differential equations
(\ref{lin1})-(\ref{lin2}) and the form of (\ref{nonlin}) depend
crucially on the values of $\alpha_i$ chosen. For the most sets of
$\alpha$'s, including the general case with all $\alpha_i\neq 0$ as
well as almost all degenerate cases with some $\alpha_i$ vanishing,
the corresponding potentials allow the separation of variables and
are ignored here.
\par The only exception to the above rule, and therefore the most
interesting quantum models, corresponds to the case\footnote{The
system with $\alpha_0 = \alpha_2 = \alpha_3 = 0;\quad \alpha_1 \ne
0$, i.e. $U = \sigma_1$ leads to analogous results with the
substitution $\chi_{\pm}\leftrightarrow\chi_{\mp}$.} when
$\alpha_0 = \alpha_1 = \alpha_2 = 0;\quad \alpha_3 \ne 0$, i.e. $U
= \sigma_3$. Then, the metric of supercharges $Q^\pm$ is
hyperbolic, i.e. $Q^{\pm}$ belong to the class discussed in
Subsection 2.2. For these models (due to (\ref{lin1}) -
(\ref{lin2}) only), the supercharges are represented in terms of
four arbitrary real functions $\mu_{1,2}$, $\mu_\pm:$ \be \chi_- =
\mu_+(x_+) + \mu_-(x_-),\qquad \chi_+  =  \mu_1(x_1) + \mu_2(x_2),
\label{chimu} \ee Hence, the last equation (\ref{nonlin})
rewritten via $\phi \equiv \mu'$ takes the form of the functional
equation: \be \phi_1(x_1)\left[ \phi_+(x_+) + \phi_-(x_-) \right]
+ \phi_2(x_2) \left[ \phi_+(x_+) - \phi_-(x_-) \right] = 0.
\label{functional} \ee It is reasonable to formulate here the
important specific property of solution (\ref{chimu}). The
superpotential \be \chi (\vec x) =\chi_+ + \chi_- = \mu_1(x_1) +
\mu_2(x_2) + \mu_+(x_+) + \mu_-(x_-); \label{property1} \ee leads
to an expression for the quantum potential $V^{(0)}(\vec x)$ (see
the first of Eq.(\ref{hams})), which also has {\bf the form of the
sum}: \be V^{(0)}(\vec x) = \Bigl(\partial_l\chi (\vec x)\Bigr)^2
- \hbar\partial_l^2\chi (\vec x) = v_1(x_1)+v_2(x_2)+v_+(x_+)
+v_-(x_-); \label{4terms1}, \ee with $v_{1,2} = \mu^{\prime\,
2}_{1,2} - \hbar \mu^{\prime\prime}_{1,2}$, $v_\pm = \mu^{\prime\,
2}_{\pm} - \hbar \mu^{\prime\prime}_{\pm}.$ It may be seen that
both terms in quantum potential (\ref{4terms1}) {\it separately}
have the form of the sums as in (\ref{property1}). Therefore, at
the quasi-classical limit $V_{cl}^{(0)}$ (see Sections 4 and 5
below), where only the first term $\Bigl(\partial_l\chi (\vec
x)\Bigr)^2$ survives, the potential is also represented in a form
similar to (\ref{4terms1}) but with truncated $v_{1,2,+,-}$. Both
in the quantum and classical contexts, form (\ref{4terms1}) seems
to be typical for a wide class of integrable two-dimensional
models, considered within very different approaches in the
literature (see \cite{perelomov}, \cite{david4}, as examples).
This is why the following statement might be useful (at least, in
the classical framework). Thus, if the general solution for the
superpotential $\chi (\vec x)$ in relation \be V_{cl}^{(0)}(\vec
x) = \Bigl(\partial_l\chi (\vec x)\Bigr)^2 =
v_1(x_1)+v_2(x_2)+v_+(x_+) +v_-(x_-) \label{relation} \ee is of
the form of (\ref{property1}), precisely the functional equation
(\ref{functional}) must be fulfilled. This equation ensures the
mutual cancellation of crossed terms in (\ref{relation}) and is
therefore very important for this class of model. The general
solution of (\ref{functional}) was found by
D.Nishnianidze\footnote{Private communication.} (see \cite{iv}):
\be {\phi}'^2_{1,2}=a\phi_{1,2}^4 + b\phi_{1,2}^2 + c;\qquad x =
\pm\int\frac{d\phi_1}{\sqrt{a\phi_1^4 + b\phi_1^2 + c}};\quad
a,b,c=const , \label{ell} \ee and explicit expressions for
$\phi_{\pm}$ can be obtained from (\ref{functional}). Among the
functions that satisfy conditions (\ref{ell}) there exists a set
of solutions of Eq.(\ref{functional}) possessing the periodicity
property. For example,
\begin{equation}
\phi_+(x) = A\frac{\text{sn}(ax|k)\,
\text{cn}(ax|k)}{\text{dn}(ax|k)};\quad \phi_-(x) =
A\frac{\text{dn}(ax|k)}{k^2\text{sn}(ax|k)\,
\text{cn}(ax|k)};\label{per}
\end{equation}
\begin{equation*}
\phi_1(x) = \phi_2(x) = B\,\,\text{dn}(\sqrt{2}ax|k),
\end{equation*}
where $A, B, a$ are real constants and $\text{sn}(\cdot|k)$,
$\text{cn}(\cdot|k)$ and $\text{dn}(\cdot|k)$ are Jacobi elliptic
functions \cite{bather} with modulus $k$. They are doubly periodic
on the complex plane of argument $x$, but in the case $k=1$ the real
period becomes infinite and the elliptic functions turn into the
hyperbolic functions $\sinh$ and $\cosh$. Restricting ourselves in
(\ref{ell}) to non-periodic functions on a whole plane $(x_1, x_2)$
satisfying (\ref{ell}), which do not happens in systems with
separation of variables, two families of models exist (see
\cite{iv}). One of them is represented by the two-dimensional {\bf
Morse} potential, with $\phi_1=\phi_2 = Be^{-\alpha x}; \phi_+ = 2A;
\phi_-=2A\coth{(\alpha x/\sqrt{2})}$:
\begin{eqnarray}
V^{(0)} =& (B^2e^{-2\alpha x_1} + \hbar B\alpha e^{-\alpha x_1}) +
(B^2e^{-2\alpha x_2} + \hbar B\alpha e^{-\alpha x_2})\nonumber\\ &
+ 2A(2A +
\hbar\frac{\alpha}{\sqrt{2}})\left[\sinh{(\frac{\alpha(x_1-x_2)}{2})}\right]^{-2}
+ 8A^2. \label{morse}
\end{eqnarray}
and the second by the two-dimensional {\bf P\"oschl-Teller}
potential with $\phi_1=-\phi_2 = A\left[ \sinh{(\sqrt{2}\alpha x)}
\right]^{-1}; \phi_+ = \phi_- =B\tanh{(\alpha x)}$:
\begin{eqnarray}
V^{(0)} &= \left( B^2 - \frac{B(B +
\hbar\alpha)}{\cosh^2{(\frac{\alpha}{\sqrt{2}}(x_1 + x_2))}}
\right) + \left( B^2 - \frac{B(B +
\hbar\alpha)}{\cosh^2{(\frac{\alpha}{\sqrt{2}}
(x_1 - x_2))}} \right) +\nonumber\\
&+A\left[ \frac{A - \hbar\sqrt{2}\alpha\cosh{(\sqrt{2}\alpha
x_1)}}{\sinh^2{(\sqrt{2}\alpha x_1)}} + \frac{A +
\hbar\sqrt{2}\alpha \cosh{(\sqrt{2}\alpha
x_2)}}{\sinh^2{(\sqrt{2}\alpha x_2)}}\right]. \label{pot1}
\end{eqnarray}
Other members of these families can be obtained by using two
discrete symmetries of solutions of Eq.(\ref{functional}): \ba
S_1:\  \biggl\{ \phi_1(x_1), \phi_2(x_2), \phi_+(x_+), \phi_-(x_-)
\biggr\} &\longrightarrow &\biggl\{\phi_+(x_1), \phi_-(x_2),
\phi_1(x_+),
\phi_2(x_-)\biggr\};\nonumber\\
S_2:\  \biggl\{\phi_1(x_1), \phi_2(x_2), \phi_+(x_+),
\phi_-(x_-)\biggr\} &\longrightarrow & \biggl\{\phi_1(x_1),
-\phi_2(x_2), \phi_+^{-1}(x_+), \phi_-^{-1}(x_-)\biggr\}
\label{S1S2} \ea and different combinations thereof.
\section*{4. Supersymmetric extensions of scalar Hamiltonians and their integrability}
\vspace*{0.3cm} \hspace*{3ex} In the previous Section we presented
the explicit forms (\ref{morse}), (\ref{pot1}) of the terms in
Eq.(\ref{4terms1}) for a certain class of quantum integrable
Hamiltonians. In this Section we shall build their classical and
quantum SUSY extensions and we shall also demonstrate their
integrability properties. \vspace*{0.2cm}
\subsection*{4.1. The classical limit for $H^{(0)}$}
\hspace*{3ex} First, we shall consider for $H^{(0)}$ its classical
limit $H^{(0)}_{cl},$ for which the integral of motion $R_{cl}$
exists: $\{H^{(0)}_{cl}, R_{cl}^{(0)} \}_P =0$ ($\{\cdot ,
\cdot\}_P$ denotes standard Poisson brackets). This can be done by
the simple limit procedure $\hbar \to 0$ in Eq.(\ref{hams}). The
practical recipe is as follows. One has to replace all {\it
operators} $-i\hbar\partial_i$ by momenta $p_i$ and skip all {\it
derivatives of functions}, which include $\hbar$ as a multiplier.
One thus obtains\footnote {Here and below we will omit the
arguments of $\phi_{1,2}$ and $\phi_{\pm}$, which implies that
$\phi_{1,2} \equiv \phi_{1,2}(x_{1,2})$ and $\phi_\pm  \equiv
\phi_\pm(x_\pm).$}:
$$
 H^{(0)}_{cl} = p_jp_j + \phi_1^2 + \phi_2^2 +
\phi_+^2 + \phi_-^2; $$ $$ Q_{cl}^ \pm = p_1^2 - p_2^2 \pm
i\sqrt{2}(\phi_+ + \phi_-)p_1 \mp i\sqrt{2}(\phi_+ - \phi_-)p_2 +
\phi_1^2  -\phi_2^2 - 2\phi_+\phi_-. $$ The integral of motion has
the form $R^{(0)}_{cl}=Q_{cl}^+Q^-_{cl}.$ Its involution with the
Hamiltonian can be checked either by direct calculation or by a
simpler two-step procedure, proposed in general form in \cite{pomi}.
It is instructive to perform it in the context of the models
considered here. First, one has to prove the intermediate relations:
\begin{equation}
\{ H^{(0)}_{cl}, Q^\pm_{cl} \}_P = \pm 2i(\phi_+' +
\phi_-')Q^\pm_{cl}. \label{interm}
\end{equation}
From the definition of Poisson brackets
\begin{eqnarray}
\{H^{(0)}_{cl}, Q^\pm_{cl} \}_P  &=&  \pm2i(\phi_+' +
\phi_-')(p_1^2 - p_2^2 \pm i\sqrt{2}(\phi_+ + \phi_-)p_1 \mp
i\sqrt{2}(\phi_+ -
\phi_-)p_2 - 2\phi_+\phi_-) \mp \nonumber\\
&\mp& 2i\sqrt{2}\left(\phi_+(\phi_1\phi_1' - \phi_2\phi_2') +
\phi_-(\phi_1\phi_1' + \phi_2\phi_2')\right).\label{interm2}
\end{eqnarray}
The last term can be transformed with the use of the functional
equation (\ref{functional}):
\begin{eqnarray}
&&\phi_+(\phi_1\phi_1' - \phi_2\phi_2') + \phi_-(\phi_1\phi_1' +
\phi_2\phi_2') = \frac{1}{\sqrt{2}}\phi_+\partial_-\left(\phi_1^2
+ \phi_2^2\right) +
\frac{1}{\sqrt{2}}\phi_-\partial_+\left(\phi_1^2 +
\phi_2^2\right) = \nonumber\\
&&=\frac{1}{\sqrt{2}}\left[ \partial_-(\phi_+(\phi_1 + \phi_2)^2)+
\partial_+(\phi_-(\phi_1 - \phi_2)^2) - 2\phi_+\partial_-
(\phi_1\phi_2) + 2\phi_-\partial_+(\phi_1\phi_2) \right] =
\nonumber\\ &&=\frac{1}{\sqrt{2}}\left[
-\partial_-(\phi_-(\phi_1^2 - \phi_2^2)) -
\partial_+(\phi_+(\phi_1^2 - \phi_2^2)) - 2\phi_+\partial_-
(\phi_1\phi_2) + 2\phi_-\partial_+(\phi_1\phi_2) \right]  =
\nonumber\\
&&=-\frac{1}{\sqrt{2}}(\phi_+' + \phi_-')(\phi_1^2 -
\phi_2^2).\nonumber
\end{eqnarray}
Substituting this into (\ref{interm2}), we prove (\ref{interm}).
Then, (\ref{interm}) leads to the involution of $R^{(0)}_{cl}$ with
$H^{(0)}_{cl}$:
$$\{ H^{(0)}_{cl}, Q^+_{cl}Q^-_{cl} \}_P = \{ H^{(0)}_{cl},
Q^+_{cl} \}_PQ^-_{cl} + Q^+_{cl}\{ H^{(0)}_{cl}, Q^-_{cl} \}_P =
0.$$ It is easy to check that the same classical limit procedure for
the second scalar Hamiltonian $H^{(2)}$ and for the components of
the matrix $H_{ik}^{(1)}$ in (\ref{hams}) leads to the simple
results: $H^{(2)}_{cl} = H^{(0)}_{cl}$ and $H^{(1)}_{ik,\, cl}=
\delta_{ik} H^{(0)}_{cl}.$ Naturally, the corresponding integrals of
motion coincide too: $R^{(2)}_{cl} = R^{(0)}_{cl}$ and
$R^{(1)}_{ik,\, cl}= \delta_{ik} R^{(0)}_{cl}.$ In the next
Subsection we shall construct another classical limit of $\hat{H}$
that will also include Grassmanian dynamical variables in addition
to $p_j$ and $x_j,$ and this can be interpreted as {\it a
SUSY-extension} of $H^{(0)}_{cl}$. \vspace*{0.2cm}
\subsection*{4.2. The SUSY extension of classical scalar Hamiltonians}
\hspace*{3ex} It is well-known \cite{abei}, \cite{wipf} that the 2D
representation of SUSY algebra, reviewed in Section 2, can be
obtained by the canonical quantization from the classical system
with the Hamiltonian:
\begin{equation}
H_{cl} = p_lp_l + (\partial_l \chi (\vec x))(\partial_l \chi (\vec
x)) + 2i\partial_i\partial_l \chi (\vec x)\psi^1_i\psi^2_l,
\label{superH}
\end{equation}
where $\psi^1_i$ and $\psi^2_j$ are Grassmanian anticommuting
variables: $\{\psi^\alpha_i, \psi^\beta_j\} = 0$ ($i,j = 1,...,d$;
$\alpha, \beta=1,2$). One can define the Poisson bracket on the
phase space of the system with classical bosonic and fermionic
variables \cite{bei} as follows:
$$\{F, G\}_P = \frac{\partial F}{\partial p_j}\frac{\partial
G}{\partial x^j} - \frac{\partial F}{\partial x^j}\frac{\partial
G}{\partial p_j} + iF\frac{\overleftarrow{\partial}}{\partial
\psi^\alpha_i}\frac{\overrightarrow{\partial}}{\partial
\psi^\alpha_i}G,$$ such that the canonical brackets are $\{p_i,
x_j\}_P=\delta_{ij};\,\,\{\psi^{\alpha}_i,
\psi^{\beta}_j\}_P=i\delta_{ij}.$ Thus, the Hamiltonian is involved
in the SUSY algebra:
$$\{ Q_\alpha, Q_\beta \}_P = i\delta_{\alpha \beta} H_{cl}; \qquad
\{ H_{cl}, Q_\alpha \}_P = 0$$ with classical supercharges
$$Q_1 = p_j\psi^1_j - (\partial_j\chi)\psi^2_j; \qquad Q_2 =
p_j\psi^2_j + (\partial_j\chi)\psi^1_j.$$ To quantize this model,
one has to introduce the bosonic operators $\hat{p}_i$ and
$\hat{x}_j$ with canonical commutation relations, and fermionic ones
$\hat{\psi}^\alpha_j$ obeying $\{ \hat{\psi}_i^\alpha,
\hat{\psi}_j^\beta\} = \hbar\delta_{ij}\delta^{\alpha \beta}$. At
this stage, it is convenient to introduce the fermionic operators
$\hat{\psi}^\pm_j = (\sqrt{2})^{-1}(\hat{\psi}^1_j \mp
i\hat{\psi}^2_j)$ with anticommutation relations $\{\hat{\psi}_i^+,
\hat{\psi}_j^-\} = \hbar\delta_{ij}$ and $\{\hat{\psi}_i^-,
\hat{\psi}_j^-\} = \{\hat{\psi}_i^+, \hat{\psi}_j^+\} = 0$. These
can be treated as creation/annihilation operators in the system of
$d$ spin $1/2$ fermions. Thus, the quantum counterpart of the
Hamiltonian (\ref{superH}) - the Superhamiltonian - in terms of
these operators takes the form:
\begin{equation}
\hat{H} = -\hbar^2\partial_j\partial_j + \left[
(\partial_j\chi)(\partial_j\chi) - \hbar(\partial_j\partial_j\chi)
\right] + 2(\partial_i\partial_j\chi)\hat{\psi}^+_i\hat{\psi}^-_j.
\label{SHquant}
\end{equation}
Together with the quantum supercharges $\hat{Q}^\pm = \pm i
(\sqrt{2})^{-1}(\hat{Q}^1 \mp i\hat{Q}^2) = (\pm i\hat{p}_j +
(\partial_j\chi))\hat{\psi}^\pm_j$ this generates the algebra
(\ref{alg}). To reproduce the matrix form (\ref{suham}) -
(\ref{hams}) one should choose the matrix representation for the
creation and annihilation operators $\hat{\psi}_j^\pm$. For $d=2$,
these operators are $4\times4$ matrices, and a possible choice is as
follows \cite{abei}: $\hat{\psi}_1^+ = \hbar^{1/2}(E_{2,1} -
E_{4,3})$, $\hat{\psi}_2^+ = \hbar^{1/2}(E_{3,1} + E_{4,2})$ and
$\hat{\psi}^-_j = (\hat{\psi}^+_j)^\dag$ (here matrices $E_{m,n}$
are defined as $(E_{m,n})_{ik}\equiv \delta_{mi}\delta_{nj}$). Thus,
we have obtained a matrix realization of two-dimensional SUSY QM
(\ref{alg}) - (\ref{hams}) by means of the canonical quantization of
a certain classical model (\ref{superH}).
\par One can see that in this representation $\hat{H}$ has
a block-diagonal structure. The origin of this feature of the model
is the conservation of the fermion number $[\hat{H}, \hat{N}] = 0$,
with fermion number operator $\hat{N} =
\hat{\psi}^+_j\hat{\psi}^-_j$. Therefore, each component of the
Superhamiltonian acts in a space of states with a fixed fermion
number.
\par In our case $d=2$, this structure is rather simple.
Let us define a basis in the state space: the vacuum $|00\!\!>$,
which is annihilated by $\hat{\psi}^-_i$, and the excited states
$|10\!\!> = \hbar^{-1/2}\hat{\psi}^+_1|00\!\!>$; $|01\!\!> =
\hbar^{-1/2}\hat{\psi}^+_2|00\!\!>$; $|11\!\!> =
\hbar^{-1}\hat{\psi}^+_2\hat{\psi}^+_1|00\!\!>$ (the
$\hbar$-dependent multipliers provide the proper normalization of
the state vectors: $<\!\!mn|mn\!\!> = 1$, $\forall m,n=0,1$). Thus
$H^{(0)}$ acts in one-dimensional space with a fermion number of $0$
and a single basis element $|00\!\!>$; for $H^{(2)}$, the fermion
number is $2$ and the basis element is $|11\!\!>$. The Hamiltonian
$H^{(1)}_{ik}$ acts in two-dimensional state space $\{ |01\!\!>,
|10\!\!> \},$ where the fermion number is $1$.
\par The conclusion to be drawn from this derivation is as
follows. Having the classical system with $H^{(0)}_{cl} =p_jp_j +
(\partial_j\chi)(\partial_j\chi)$, one can construct its SUSY
extension of the form (\ref{superH}). This classical SUSY extension
can be quantized canonically to obtain the quantum Superhamiltonian
(\ref{SHquant}), the original $H^{(0)}_{cl}$ being the classical
limit of $H^{(0)}$ - the first scalar component of the quantum
Superhamiltonian. In our case, $H^{(0)}_{cl}$ was integrable (see
previous Subsection), and we shall explicitly find the integral of
motion $R_{cl}$ for its quantum SUSY-extension (Superhamiltonian).
An analogous problem was investigated by alternative methods in
\cite{Aai}, but for a much more narrow class of models (amenable to
separation of variables). \vspace*{0.2cm}
\subsection*{4.3. Integrals of motion for the quantum and classical SUSY extensions}
\hspace*{3ex} We start from construction of the quantum integral of
motion $\hat{R}$ for the Superhamiltonian (\ref{SHquant}):
$[\hat{H}, \hat{R}] = 0,$ i.e. of conserved operators $R^{(i)}$ for
each component of the Superhamiltonian:
\begin{equation}
[H^{(0)}, R^{(0)}] = 0;\quad [H^{(2)}, R^{(2)}] = 0; \label{hr02}
\end{equation}
\begin{equation}
[H^{(1)}, R^{(1)}] = 0;\label{hr1}
\end{equation}
(note that the last commutator is the matrix one). The explicit
expression for $R^{(0)}$ can be obtained from (\ref{symm}) and
(\ref{twist}):
\begin{equation}
R^{(0)} =
q^+_iU_{ik}\tilde{q}^-_k\tilde{q}^+_mU_{ml}q^-_l.\label{R0}
\end{equation}
One can obtain $R^{(2)}$ from $R^{(0)}$ by the substitutions
$q^\pm_j \to -q^\mp_j$ and $\tilde{q}^\pm_j \to -\tilde{q}^\mp_j$
(since $H^{(0)}$ turns to $H^{(2)}$ after the substitution
$\chi(\vec{x}) \to -\chi(\vec{x})$):
\begin{equation}
R^{(2)} = q^-_iU_{ik}\tilde{q}^+_k\tilde{q}^-_mU_{ml}q^+_l.
\label{R2}
\end{equation}
With respect to $R^{(1)}$, the form of intertwining relations
(\ref{2dintw}) tells us how to build this symmetry operator
explicitly. One can check that the following matrix operator {\it
of sixth order in derivatives}
\begin{equation}
R^{(1)}_{ik} = q^-_iR^{(0)}q^+_k, \label{R1}
\end{equation}
satisfies Eq.(\ref{hr1}).
\par It is clear from the material of the previous Subsection
that knowledge of $R^{(i)},\,\, (i=0,1,2)$ provides a symmetry
operator $\hat R$ for the whole block-diagonal Superhamiltonian
$\hat H.$ In order to construct an all-sector expression for $\hat
R:$
\begin{equation}
\hat{R} = R^{(0)}P^{(0)} + R^{(2)}P^{(2)} +
R^{(1)}_{ik}P^{(1)}_{ik}, \label{decomp}
\end{equation}
we introduce the corresponding projection operators
$P^{(i)}\,\,(i=0,1,2)$ onto subspaces with definite fermion numbers.
The scalar projectors $P^{(0)}$ and $P^{(2)}$ give unity when acting
on $|00\!\!>$ and $|11\!\!>,$ respectively, and give zero otherwise.
The components $P^{(1)}_{ik} \,(i,k=1,2)$ of the $2\times 2$ matrix
projector operators $P^{(1)}$ transform the $k-$th component of the
state vector into its $i-$th component\footnote{By definition, "the
first component" is $|10\!\!>,$ and "the second component" is
$|01\!\!>$.} and are zero on other components of the state vector.
One can check directly that an explicit form of these operators $P$
leads to:
\begin{eqnarray}
\hat{R} &=&
-R^{(0)}\hbar^{-2}\hat{\psi}_1^-\hat{\psi}_2^-\hat{\psi}_1^+\hat{\psi}_2^+
-
R^{(2)}\hbar^{-2}\hat{\psi}_1^+\hat{\psi}_2^+\hat{\psi}_1^-\hat{\psi}_2^-
-
R^{(1)}_{11}\hbar^{-2}\hat{\psi}_1^+\hat{\psi}_2^-\hat{\psi}_1^-\hat{\psi}_2^+
- \nonumber \\ & & -
R^{(1)}_{22}\hbar^{-2}\hat{\psi}_1^-\hat{\psi}_2^+\hat{\psi}_1^+\hat{\psi}_2^-
+ R^{(1)}_{12}\hbar^{-1}\hat{\psi}_1^+\hat{\psi}_2^- +
R^{(1)}_{21}\hbar^{-1}\hat{\psi}_2^+\hat{\psi}_1^-, \label{hatr}
\end{eqnarray}
where $R^{(i)}$ are given by (\ref{R0}) - (\ref{R1}). Naturally,
(\ref{hatr}) can be simplified by employing anticommutation
relations for $\hat{\psi}^\pm_i$. One should not be confused by the
presence of the negative powers of the Planck constant in
Eq.(\ref{hatr}) since they disappear in all matrix elements for
$\hat{R}$.
\par Let us prove straightforwardly that the $\hat{R}$ constructed commutes
with $\hat{H}$. The Superhamiltonian can be presented similarly to
(\ref{decomp}):
$$\hat{H} = H^{(0)}P^{(0)} + H^{(2)}P^{(2)} +
H^{(1)}_{ik}P^{(1)}_{ik},$$ where $H^{(i)}$ are given by
Eqs.(\ref{hams}). By definition $[P^{(0)}, P^{(2)}] = [P^{(1)}_{ik},
P^{(2)}] = [P^{(1)}_{ik}, P^{(0)}] = 0$, and therefore, due to
(\ref{hr02}), we have:
$$
[\hat{H}, \hat{R}] = [H^{(1)}_{ij}P^{(1)}_{ij},
R^{(1)}_{kl}P^{(1)}_{kl}].
$$
Employing the explicit form (\ref{hatr}) of $P^{(i)}$ and
Eq.(\ref{hr1})), one can see that the commutator in the rhs
vanishes, completing our proof.
\par From expression (\ref{hatr}), its classical limit $R_{cl}$
can be constructed by means of the substitution
$\hbar^{-1/2}\hat{\psi}^\alpha_j \to \psi^\alpha_j$ for fermionic
operators. Finally, the total classical integral of motion reads:
\begin{eqnarray}
R_{cl} & = & R^{(0)}_{cl} + \frac{i}{2}(R^{(1)}_{11\,cl} -
R^{(1)}_{22\,cl})(\psi^1_1\psi^2_1 - \psi^1_2\psi_2^2) +
\frac{1}{2}(R^{(1)}_{21\,cl} - R^{(1)}_{12\,cl})(\psi_1^1\psi_2^1
+ \psi_1^2\psi_2^2) +\nonumber \\ && +
\frac{i}{2}(R^{(1)}_{12\,cl} + R^{(1)}_{21\,cl})(\psi_1^2\psi_2^1
+ \psi_1^1\psi_2^2) +
(R_{11\,cl}^{(1)}+R_{22\,cl}^{(1)}-2R^{(0)}_{cl})\psi_1^1\psi_1^2\psi_2^1\psi^2_2,
\label{Rfinal}
\end{eqnarray}
where its components are:
\begin{eqnarray}
R^{(0)}_{cl} & = & R^{(2)}_{cl} = \left( p_1^2 - p_2^2 + \phi_1^2
- \phi_2^2 - 2\phi_+\phi_- \right)^2 + 2\left( (\phi_+ +
\phi_-)p_1 - (\phi_+
- \phi_-)p_2 \right)^2; \nonumber\\
R^{(1)}_{11\,cl} & = & \left(p_1^2 + \phi_1^2 +
\frac{1}{2}(\phi_+^2 +
\phi_-^2) + 2\sqrt{2}\phi_+\phi_- \right)R^{(0)}_{cl}; \nonumber\\
R^{(1)}_{22\,cl} & = & \left(p_2^2 + \phi_2^2 +
\frac{1}{2}(\phi_+^2 -
\phi_-^2) - 2\sqrt{2}\phi_+\phi_- \right)R^{(0)}_{cl}; \label{iint}\\
R^{(1)}_{12\,cl} & = & \left( ip_1 + \phi_1 +
\frac{1}{\sqrt{2}}(\phi_+ + \phi_-) \right) \left( -ip_2 + \phi_2
+ \frac{1}{\sqrt{2}}(\phi_+ - \phi_-) \right)R^{(0)}_{cl}; \nonumber\\
R^{(1)}_{21\,cl} & = & \left( ip_2 + \phi_2 +
\frac{1}{\sqrt{2}}(\phi_+ - \phi_-) \right) \left( -ip_1 + \phi_1
+ \frac{1}{\sqrt{2}}(\phi_+ + \phi_-) \right)R^{(0)}_{cl}.
\nonumber
\end{eqnarray}
\vspace*{0.2cm}
\section*{5. The flipped potentials and the classical Hamilton-Jacobi equation}
\vspace*{0.5cm} \hspace*{3ex} In this Section we shall establish
links between the Hamilton-Jacobi equations of Classical Mechanics
and the equation for the superpotential. Starting from
Eq.(\ref{relation}), one can see that the condition for the
classical Hamiltonian
$$H_{cl} = p_l^2 + V(\vec{x})
$$
to be supersymmetric with superpotential $\chi(\vec{x})$ takes the
form:
\begin{equation}
V(\vec{x}) = \frac{\partial\chi(\vec{x})}{\partial
x_l}\frac{\partial\chi(\vec{x})}{\partial x_l}. \label{ham}
\end{equation}
On the other hand, for the Hamiltonian $h_{cl}=p_l^2+U(\vec x)$ with
potential $U(\vec x)$ and the classical action functional $S$, the
well known Hamilton-Jacobi equation \cite{hamilton} reads:
\begin{equation}
 \frac{\partial S}
{\partial t} + h_{cl}(\frac{\partial S} {\partial
x_1},\frac{\partial S}{\partial x_2}, \cdots \frac{\partial S}
{\partial x_d}; x_1,x_2,\cdots ,x_d)=0 \label{HJ}.
\end{equation}
There being no explicit dependence on time in $h_{cl}$, one looks
for its solutions of the form $S(t,x_1,x_2,\cdots
,x_d)=W(x_1,x_2,\cdots ,x_d)-Et$, and the time-independent
Hamilton-Jacobi equation becomes:
\begin{equation}
E=(\partial_lW)^2+U(\vec{x}), \label{eq:thj}
\end{equation}
where $W(x_1,x_2,\cdots ,x_d)$ is usually referred to as the
Hamilton characteristic function. Solutions of (\ref{eq:thj}) in the
case $E=0$ are obviously connected with those of (\ref{ham}):
\begin{equation}
\chi(\vec{x}) = \pm iW(\vec{x}). \label{conn}
\end{equation}
Eq.(\ref{eq:thj}), with zero energy, can alternatively be thought of
as a condition for the ''flipped'' classical potential $V = -U$ to
be supersymmetric, i.e. $V$ should satisfy (\ref{ham}), with $\chi$
and $W$ related by (\ref{conn}). Thus, we find
\[
W(x_1,x_2)=\mp
i\left[\mu_1(x_1)+\mu_2(x_2)+\mu_+(x_+)+\mu_-(x_-)\right]
\]
as the Hamilton characteristic function of the system. The system of
equations of motion for $E=0$ \ba \dot{x}_1&=&\frac{\partial
W}{\partial x_1}=\mp
i\left[\phi_1+{1\over\sqrt{2}}\phi_++{1\over\sqrt{2}}\phi_-\right]\nonumber
\\ \dot{x}_2&=&\frac{\partial W}{\partial x_2}=\mp
i\left[\phi_2+{1\over\sqrt{2}}\phi_+-{1\over\sqrt{2}}\phi_-\right]\label{mot}
\ea is not amenable to separation of variables and in general has
(non-physical) complex solutions. One may see that system
(\ref{mot}) becomes real, and bona fide solutions exist for the
specific complexification of the P\"oschl-Teller model (\ref{pot1}):
namely, with purely imaginary $\alpha$. In contrast to the
complexification of two-dimensional Morse potential in
\cite{pseudo}, this one is PT-invariant.
\par Analogous classical systems with "flipped" potentials
were investigated in \cite{Aai} for the case of $d=2$ integrable
models of the Liouville type. For these systems the Hamilton-Jacobi
equations were separable in elliptic, polar, parabolic and Cartesian
coordinates. The structure of related supersymmetric models (also
with separation of variables) {\bf in the quantum domain} has been
investigated in \cite{Aai1} via canonical quantization. In
particular, it was shown that there are two essentially different
supersymmetric extensions (two different superpotentials) for a
given separable classical solution of the Hamilton-Jacobi equation.
In the present paper our strategy is just the opposite. Namely,
starting from scalar quantum $d=2$ systems which {\bf do not allow}
for separation of variables, but do have non-trivial symmetry
operators, we construct their quantum SUSY-extension. Then, we
describe corresponding classical SUSY-extended systems and their
integrals of motion. Finally, the link between this kind of
classical system and the Hamilton-Jacobi approach for related
systems with "flipped" potential provides the integrability of these
"flipped" systems too. The necessary explicit expressions for
integrals of motion can be obtained from (\ref{Rfinal}),
(\ref{iint}) by the substitution $\phi \rightarrow \pm i\phi$, which
is equivalent to (\ref{conn}). It should be remarked that
(analogously to \cite{Aai1}) besides an arbitrary common sign in
(\ref{conn}) there is the additional non-uniqueness of the
superpotential for this class of model. Indeed, equation (\ref{ham})
has {\it two independent} solutions for fixed original classical
potential: $\chi (\vec x)$ and $\tilde\chi (\vec x).$ To prove this
statement, one has to check that $(\partial_l\chi)^2 =
(\partial_l\tilde\chi)^2$
 for $\chi = \mu_1+\mu_2+\mu_++\mu_-$
and $\tilde\chi = \mu_1+\mu_2-\mu_+-\mu_-$ (see (\ref{chimu})) due
to the nonlinear equation (\ref{nonlin}). \vspace*{0.2cm}
\section*{Acknowledgements}
\vspace*{0.5cm} \hspace*{3ex} The work was partially supported by
the Spanish MEC grants SAB2004-0143 (M.V.I.), MTM2005-09183
(M.V.I.), BFM2003-00936 (J.M.G.), by the project VA013C05 of the
Junta de Castilla y Le\'on (J.M.G. and M.V.I.) and by Russian
grants (M.V.I. and P.A.V.) RFFI 06-01-00186-a and RNP 2.1.1.1112.
P.A.V. is indebted to the International Centre of Fundamental
Physics in Moscow, the non-profit foundation "Dynasty" and the
Grant Centre of Natural Sciences (grant M05-2.4D-56) for financial
support. M.V.I. is grateful to the University of Salamanca and the
University of Valladolid for kind hospitality and to J.Negro and
L.M.Nieto for helpful discussions.

\end{document}